\documentclass[aps,prd,twocolumn,groupedaddress,nofootinbib]{revtex4-1}
\usepackage[utf8]{inputenc}
\usepackage[english]{babel}
\usepackage[T1]{fontenc}
\usepackage{amsmath,amsfonts,amsthm,bm}
\usepackage{amsfonts}
\usepackage{hyperref}
\usepackage{amssymb}
\usepackage{graphicx}
\usepackage{physics}
\usepackage{array}
\usepackage{gensymb}
\usepackage[normalem]{ulem}
\usepackage{aas_macros}

\newcommand{\hie}{hierarchical }

\begin{document}

%\preprint{}

\title{Precession resonances in hierarchical triple systems}


\author{Adrien Kuntz}
\email[]{adrien.kuntz@sns.it}
%\thanks{}
\affiliation{Scuola Normale Superiore, Piazza dei Cavalieri 7, 56126, Pisa, Italy}
\affiliation{INFN Sezione di Pisa, Largo Pontecorvo 3, 56127 Pisa}


\date{\today}

\begin{abstract}
We describe a new kind of resonance occuring in relativistic three-body hierarchical systems: the precession resonance, occuring when the relativistic precession timescale of a binary equals the period of a distant perturber. We find that, contrary to what most previous studies assume, it can lead to an exponential increase of eccentricity of the binary even when relativistic precession dominates the quadrupolar perturbation. The resonance may happen in the observation band of LISA or change the eccentricity distribution of triples. We discuss the physics of the resonance, showing that it mainly depends on three parameters.
\end{abstract}

%\pacs{}
%\keywords{}

\maketitle

\section{Introduction}

Resonances in many-body systems, occuring when at least two different frequencies take commensurate values, have played a major role in astronomy ever since the first studies of Poincaré and Laplace. Already in Newtonian mechanics, the phenomenology of orbital resonances is quite rich since they can lead to either stable orbits (this is the case of Pluto~\cite{1971AJ.....76..167W} or the Trojans~\cite{1993CeMDA..57...59M}) or unstable ones (the Kirkwood gaps in the asteroid belt~\cite{1982AJ.....87..577W} or the Cassini division in Saturn's rings~\cite{1978Icar...34..240G}). It is not surprising that relativistic effects bring several new interesting features in resonant systems which are not present in Newtonian mechanics. For example, even a two-body system may display the so-called "transient resonances"~\cite{2012PhRvL.109g1102F} which is a unique feature of relativistic effects, whereas three-body systems can undergo new kinds of resonances generalizing the Newtonian case~\cite{Bonga_2019,Gupta:2021cno,Yang_2019} or present purely GR effects such as light-ring resonances~\cite{2019, cardoso2021gravitational}. 

Since three-body systems are quite common in our Universe~\cite{Martinez_2020, O_Leary_2016, Tokovinin_2006, Robson_2018}, relativistic resonances may be of importance in the advent of Gravitational Waves (GWs) astronomy, either through a direct imprint of the resonance in the gravitational waveform of black holes binaries (BHB)~\cite{Bonga_2019, Yang_2019,2015PhRvL.114h1102B, Gupta:2019unn, Chandramouli:2021kts, Gupta:2021cno, 2014PhRvD..89h4028F}, or by the modification of the original distribution of eccentricity of merging BHB~\cite{2018ApJ...856..140H,Nishizawa:2016eza,10.1093/mnras/stu039}. This last effect is known to occur in the Kozai-Lidov (KL) resonances~\cite{1962AJ.....67..591K,LIDOV1962719} which can induce large eccentricity oscillations in a (so-called inner) binary system if it is orbited by a highly inclined (so-called outer) perturber.

It is generally admitted that post-Newtonian (PN) relativistic effects quench the KL resonance because they induce a supplementary perihelion precession which dephases the resonant frequencies~\cite{Liu_2014, Naoz_2016, 2014ApJ...781L...5D}. However, if relativistic effects are strong, the perihelion precession timescale itself could be comparable to the period of the outer perturber, leading to a resonant behavior. The purpose of this article is to study the effects of such kind of resonance, which we term by "precession resonance". Note that it is a phenomenon distinct from the "tidal resonances" recently unveiled in~\cite{Bonga_2019,Gupta:2021cno}, or from another kind of precession resonance where the precession timescales of \textit{both} inner and outer orbits are commensurate, discussed in~\cite{Liu_2020, Liu:2015wgi}.
 Although the consequences of a similar type of resonance have already been studied in the Solar system (the $\nu_6$ resonance between asteroids and Saturn~\cite{1986A&A...166..326F, 1999ssd..book.....M}, however in this case the precession is not due to GR effects), we are not aware of any previous study of the impact of precession resonances on relativistic three-body systems. We will show that precession resonances can induce an exponential growth of the eccentricity of the inner binary even in regions where the KL timescale is much greater than the relativistic precession timescale, thus invalidating previous claims in the literature. This could affect gravitational wave observables through both the channels mentioned before, i.e. by modifying the eccentricity distribution of BHB or by directy affecting their gravitational waveform observable in low-frequency GW detectors like the future space-borne interferometer LISA~\cite{2013GWN.....6....4A}.

\section{Resonant hierarchical systems}

We consider a three-body system in a hierarchical configuration where an inner binary, with masses $m_1$ and $m_2$, is orbited by a distant perturber $m_3$.
We can decompose the motion into two osculating ellipses called inner (resp outer) orbit, of frequency $n$ (resp. $n_3$), with planetary elements $a,e,\omega,\iota,\Omega,u$ (resp. $a_3,e_3,\omega_3,\iota_3,\Omega_3,u_3$) consisting in the semimajor axis, eccentricity, argument of perihelion, inclination, longitude of ascending node, and true anomaly of the orbit. The outer orbit is built by replacing the inner binary with an effective point-particle located at its center-of-mass, as explained in~\cite{Kuntz:2021ohi}. The \hie assumption only assumes $a \ll a_3$, so that these elements evolve solwly in time due to the interactions between the two orbits  We will always be interested in dynamics on timescales greater than the period of the inner binary, so we will average all quantites over one orbit of the inner binary; however, it will be crucial not to average over one outer orbit in order to account for the effect of precession resonances. The Hamiltonian of the three-body system can then be written as~\cite{Kuntz:2021ohi,Naoz_2016}
\begin{equation} \label{eq:Hamiltonian}
\mathcal{H} = - \frac{G_N m \mu}{2 a} - \frac{G_N M \mu_3}{2 a_3} - 3 \mu \frac{G_N^2 m^2}{a^2 \sqrt{1-e^2}} + \mathcal{H}_\mathrm{quad}
\end{equation}
%- 3 \mu_3 \frac{G_N^2 M^2}{a_3^2 \sqrt{1-e_3^2}}
where $m=m_1+m_2$ and $\mu = m_1 m_2/m$ (resp. $M=m_1+m_2+m_3$ and $\mu_3 = m m_3/M$) are the total and reduced mass of the inner (resp. outer) binary, and $G_N$ is Newton's constant. The first two terms in this Hamiltonian corresponds to the Newtonian energies of the two orbits. Since $a$ is conserved by virtue of averaging the Hamiltonian over one inner orbit~\cite{Kuntz:2021ohi,Naoz_2016}, we could as well drop the first term. However, $a_3$ may be free to vary. The third term corresponds to the Hamiltonian inducing relativistic precession of the inner orbit. Finally, the last term is the Newtonian quadrupolar coupling between the two orbits, given in Appendix~\ref{sec:quad}.
Before moving on, let us emphasize that in the Hamiltonian~\ref{eq:Hamiltonian}, we have neglected relativistic effects on the outer orbit, as well as the relativistic coupling between the angular momentums of the two orbits. In the power-counting rules presented in~\cite{Kuntz:2021ohi}, this corresponds to an accuracy $v^2$ and $\varepsilon^2$ while dropping terms scaling as $v^2 \varepsilon$ and higher, where $v^2 = G_N m/a$ and $\varepsilon = a/a_3$. As we prove in Appendix~\ref{sec:higher}, this accuracy is sufficient to quantitatively describe precession resonances. 

The equations of motion stemming from the Hamiltonian~\eqref{eq:Hamiltonian}, and which describe the evolution of planetary elements of both orbits on long timescales, are called the Lagrange Planetary Equations (LPE) and they are given in Appendix~\ref{sec:quad}.
It is quite simple to obtain the time-evolution of the system using a numerical integrator. This evolution is generically characterized by two timescales corresponding respectively to the third and fourth term in the Hamiltonian~\eqref{eq:Hamiltonian}:
\begin{align}
t_\mathrm{PN} &= \frac{a}{3} \bigg( \frac{a}{G_N m} \bigg)^{3/2} \; , \\
t_\mathrm{KL} &= \frac{a_3^3 m^{1/2}}{G_N^{1/2} a^{3/2} m_3}
\end{align}
On a post-Newtonian timescale $t_\mathrm{PN}$, the perihelion precesses by an order-one quantity following the equation
\begin{equation} \label{eq:omegadotPN}
\dot \omega_\mathrm{PN} = \frac{1}{t_\mathrm{PN} (1-e^2)}
\end{equation}
which originates from the third term in the Hamiltonian~\eqref{eq:Hamiltonian}.
The second (Kozai-Lidov) timescale, stemming from the quadrupolar coupling between the two orbits, rules out the dynamics of all other osculating elements on top of inducing a supplementary perihelion precession not written in Eq.~\eqref{eq:omegadotPN}. %(see Refs.~\cite{10.1093/mnras/stt302,Naoz_2016} for the expression of time-derivative of osculating elements in the case where the Hamiltonian is averaged over one outer orbit).
%while all the other osculating parameters vary on the Kozai-Lidov timescale $t_\mathrm{KL}$. 
The usual treatment of such hierarchical systems, following the pioneering work of Kozai and Lidov~\cite{1962AJ.....67..591K,LIDOV1962719}, supposes that both $t_\mathrm{PN}$ and $t_\mathrm{KL}$ are much longer than the period of the outer orbit. One can then get simplified equations by averaging the Hamiltonian over one period of the outer binary~\cite{10.1093/mnras/stt302,Naoz_2016}. The dynamics is then dictated by which timescale dominates the evolution:
\begin{itemize}
\item $t_\mathrm{PN} \gg t_\mathrm{KL}$: The quadrupolar force is of greater magnitude than PN effects. The inner system is characterized by KL oscillations of eccentricity and inclination whose origin is in an exchange of angular momentum between the two orbits.
\item $t_\mathrm{PN} \sim t_\mathrm{KL}$: The quadrupolar force has an equivalent strength to PN effects. Several interesting behaviors can emerge from the non-trivial interplay between quadrupolar and PN terms~\cite{Naoz_2013, 2019ApJ...883L...7L, Liu_2020, Fang_2019, Fang_2020, PhysRevD.89.044043, PhysRevLett.120.191101, Lim:2020cvm, Migaszewski:2008tp}
\item $t_\mathrm{PN} \ll t_\mathrm{KL}$: PN terms dominate the evolution. This suppresses greatly the magnitude of KL oscillations, so that the outer orbit cannot induce sizeable changes in the eccentricity of the inner orbit~\cite{Liu_2014, Naoz_2016, 2014ApJ...781L...5D}.  
\end{itemize}
In some cases, the KL mechanism can induce so great eccentricities that it can make a binary system merge within one KL oscillation~\cite{Randall:2018nud, 10.1093/mnras/stu039, 2018ApJ...856..140H}. However, if the inclination of the system is not large enough, the quadrupolar KL mechanism is uneffective at producing large eccentricites and the inner binary can be brought to merger only by radiation-reaction forces. As soon as it enter the PN-dominated zone, $t_\mathrm{PN} \ll t_\mathrm{KL}$, it will behave very similarly to an isolated binary system whose eccentricity evolves towards negligible values due to the effect of radiation-reaction. In this picture, then, there is little hope of measuring a nonzero eccentricity in the waveform of the inner binary.

The purpose of this article is to show that a particular type of resonance can completely alter the time-evolution of the eccentricity in the PN dominated domain $t_\mathrm{PN} \ll t_\mathrm{KL}$. Indeed, if the inner binary is relativistic enough, one can imagine a situation where its precession frequency is in phase with the (inverse) period of the outer orbit. Thus, let us define a precession resonance by the following condition:
\begin{equation} \label{eq:resonance_cond}
q \dot \omega = p n_3 \; ,
\end{equation}
where $p,q$ are integers.
This mean that the \textit{perihelion} of the inner orbit should complete $p$ revolutions when the third object completes $q$ revolutions. By its very essence, such a resonance requires that the inner binary should be quite relativistic; furthermore, this also means that the PN timescale $t_\mathrm{PN}$ should be of the same order-of-magnitude as the outer binary period. Consequently, it is no longer sensible to average the Hamiltonian over one outer orbit, because we are not in the approximation that $t_\mathrm{PN}$ is a long timescale. This is why the standard treatment of the post-Newtonian KL mechanism was unable to predict the effect of these resonances, so that to our knowledge they have never been reported in the literature up to now (for other examples of non-secular effects, see e.g~\cite{2013PhRvL.111f1106S, Luo_2016, PhysRevD.89.044043,Will:2020tri, Naoz_2016, 10.1093/mnras/stu039, 2014MNRAS.438..573B, 2014ApJ...781...45A}). Notice that a similar kind of resonance can occur in the Solar system (the $\nu_6$ resonance in the asteroid belt~\cite{1986A&A...166..326F, 1999ssd..book.....M}), however in this case the perihelion precession is not due to relativistic effects.

%One can see that the precession resonances can alter the evolution of the system by using the following argument. By studying the Hamiltonian, 
%The usual procedure for describing the system on timescales longer than the period of the outer orbit is to carry out an average of the Hamiltonian over one cycle of the outer orbit CITE. However, this completely neglects the resonance effect that we are going to study. Indeed, 
Where are resonant terms hidden in the Hamiltonian displayed in~\eqref{eq:Hamiltonian} ? By expanding the eccentric anomaly as a Fourier series in time, one can see that
the quadrupolar coupling in Eq.~\ref{eq:Hquad} contains terms oscillating as e.g. $\cos (2 \omega -  p n_3 t)$, where $p \in \mathbb{Z}$.
 %Expanding the mean anomaly as a Fourier series in time, we can convert such terms to a sum of terms scaling as e.g $\cos (2 \omega - 2 p \pi t/P_3)$, where $p \in \mathbb{Z}$.
Far from precession resonances, one can average this term over one period of the outer orbit, yielding a zero answer apart from $p=0$. However, close to a resonance this term behaves as a constant and will not vanish. The resonance condition is exactly the one written in~\eqref{eq:resonance_cond}, with $q=2$ given by our quadrupolar approximation. Had we included the octupole, there would also have been $q=3$ resonances. Physically, the presence of resonant terms in the Hamiltonian~\eqref{eq:Hamiltonian} makes possible an exchange of energy between the inner and outer orbits. More precisely, the outer orbit can transfer its Newtonian energy (second term in the Hamiltonian~\eqref{eq:Hamiltonian}) to the PN energy of the inner orbit (third term in the Hamiltonian~\eqref{eq:Hamiltonian}), leading to a drastic change in the inner orbit eccentricity as we will see later on.
%TODO : shall I talk of dipole resonances ? A: NO

\begin{figure}
\includegraphics[width=\columnwidth]{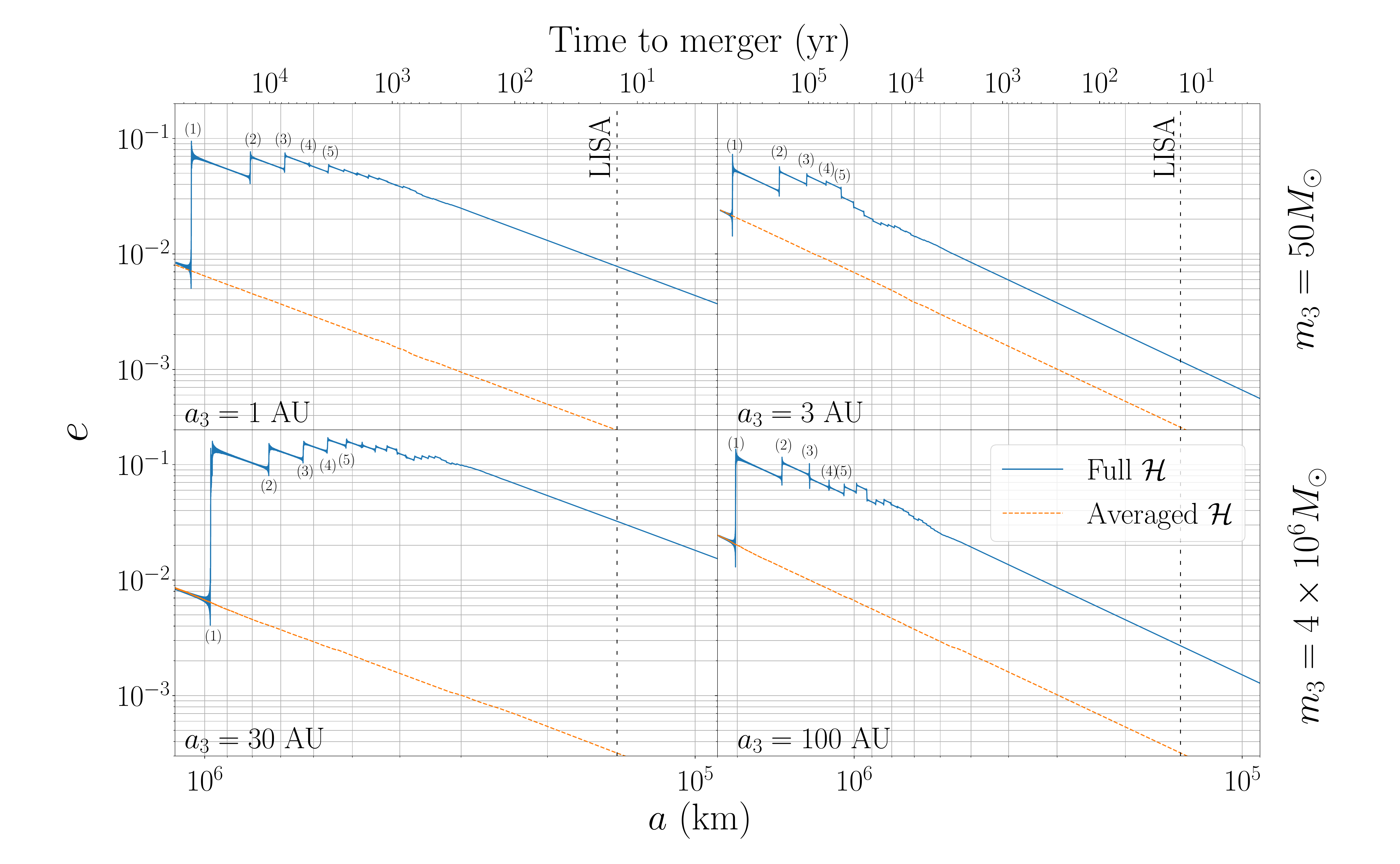}
\caption{\textit{Effect of the resonances on the evolution of the eccentricity of the inner binary}: We consider an inner BHB system of total mass $m=50 M_\odot$ and mass ratio $\nu=m_1 m_2/m^2=0.15$. The semimajor axis $a$ is initialized to $a=0.02$AU with an eccentricity of $e=0.03$, corresponding to the expected radiation-reaction decrease from a moderate $e$ a Hubble time ago. The mass of the BH perturber is either $m_3=50 M_\odot$ (upper panels) or $m_3= 4 \times 10^6 M_\odot$ (lower panels), and its distance is varied from $1$AU (top left) to $100$AU (bottom right).
The other initial conditions of the system are $e_3=0.7$, $\omega=\omega_3=\Omega=0 \degree$ and $\iota_\mathrm{tot} = \iota + \iota_3 = 60 \degree$.
The continuous blue curve corresponds to the eccentricity of the inner binary $e$ obtained by solving the LPE with the complete Hamiltonian~\eqref{eq:Hamiltonian}, which is not averaged over the period of the outer orbit. For comparison, we show in dashed orange the evolution of the eccentricity obtained by averaging the Hamiltonian~\eqref{eq:Hamiltonian}. The vertical black dashed line corresponds to a GW frequency of $f_\mathrm{GW}=15$mHz in the LISA band~\cite{LISAscience,2013GWN.....6....4A}.The number close to each resonance denote the corresponding value of $p$ in Eq.~\eqref{eq:resonance_cond} with $q=2$;
note that higher $p$'s correspond to smaller resonance effects.
}
\label{fig1}
\end{figure}

A resonance condition may appear as a fine-tuning between two incommensurable frequencies, namely the inner precession timescale and the outer period. However, if the inner binary evolves under the effect of radiation-reaction, its orbit will shrink so that its precession timescale will shorten. Thus, starting from an initial binary precessing too slowly to be in resonance with the distant perturber, radiation-reaction will induce the inner binary to pass through \textit{all} the resonances conditions before merger. It is therefore important to study the effect of precession resonances for an inner binary driven by radiation-reaction. To this aim, we numerically solve the system of differential equations from the LPE~\eqref{eq:dot_a}-\eqref{eq:dot_Omega} for both inner and outer orbit, augmented with the radiation-reaction (RR) terms for the inner orbit~\cite{PhysRev.131.435, Maggiore:1900zz}:
\begin{align}
\left. \frac{\mathrm{d}a}{\mathrm{d}t} \right\vert_\mathrm{RR} &= - \frac{64}{5} \frac{G_N^3 m^3 \nu}{a^3 (1-e^2)^{7/2}} \bigg(1 + \frac{73}{24}e^2 + \frac{37}{96} e^4 \bigg) \; , \\
\left. \frac{\mathrm{d}e}{\mathrm{d}t} \right\vert_\mathrm{RR} &= - \frac{304}{15} \frac{G_N^3 m^3 \nu}{a^4 (1-e^2)^{5/2}} \bigg(1 + \frac{121}{304}e^2 \bigg) \; ,
\end{align}
where $\nu = \mu/m$ is the symmetric mass ratio of the inner binary, and the evolution of other planetary elements is left unchanged by radiation-reaction. Figure~\ref{fig1} shows the huge impact that resonances can have on the evolution of the inner binary in the PN dominated zone $t_\mathrm{PN} \ll t_\mathrm{KL}$.
 %with $m=50 M_\odot$ (the other parameters are given in the Figure).
When the resonance condition is met, the eccentricity of the inner binary increases exponentially in a very short time up to values which can reach $e \simeq 0.1$ even starting from negligible values. For comparison, we also show the eccentricity computed using the Hamltonian~\eqref{eq:Hamiltonian} averaged over one orbit of the outer binary, which cannot account for the effect of resonances as discussed previously. In this case, the evolution of $e$ is virtually indistinguishible from the power-law decrease $e/e_0 \simeq (a/a_0)^{19/12}$ expected from the inspiral of an isolated binary~\cite{Maggiore:1900zz}. This happens because these systems are very relativistic, so that the inner binary precession completely suppresses KL oscillations. Furthermore, our integrations are performed with relative inclination of the two orbits $\iota_\mathrm{tot} = \iota + \iota_3 = 60 \degree$, which means that, even if the inner binary was less relativistic, the KL mechanism in the quadrupole approximation would not induce extreme eccentricites.

All of the systems displayed in Figure~\ref{fig1} would enter the LISA bandwidth at a GW frequency of $f_\mathrm{GW} = 15$mHz with eccentricities greater than $10^{-3}$, which is potentially measurable given the planned sensitivity~\cite{LISAscience}. Not taking into account the effect of resonances would lead to the incorrect estimate of $e$ being beyond the detection limit.
%Not taking into account the effect of resonances would lead to the false estimate that $e \simeq 3 \times 10^{-4}$ at $f_\mathrm{GW} = 15$mHz, which is below the detection limit.
Furthermore, note that the range of parameters considered in Fig~\ref{fig1} is not unrealistic for three-body systems, although it requires a quite small $a_3$. BH triples of similar mass formed in globular clusters often feature outer semi-major axis of a few AU~\cite{Antonini_2016, 2014ApJ...793..137N, 2015ApJ...800....9M, 2013MNRAS.430.2262H, 2002, 2006}, while mass segregation can bring binaries at short distances from supermassive BHs (SMBHs) $a_3 < 100$ AU ~\cite{2009, Sari:2019hot, 2019MNRAS.488.4370F, Zhang:2021pwe}.

On the other hand, several other mechanisms can lead to non-negligible eccentricites in the LISA bandwidth, including the KL mechanism for binaries with nearly perpendicular inclinations~\cite{10.1093/mnras/stu039,2018ApJ...856..140H, Naoz_2013,Naoz_2016,Seto:2013wwa,Stephan_2016, Randall:2018nud, 2018, Samsing:2018isx, DOrazio:2018jnv} (as previously stated, the KL oscillations would take place only when $t_\mathrm{PN} \gtrsim t_\mathrm{KL}$, i.e for separations of the inner binary wider than the ones considered in Figure~\ref{fig1}). Disentangling the effect of precession resonances from others would necessitate a detailed population study, which is beyond the scope of this article.  
 A clear-cut identification of a precession resonance could be possible if one of the resonance lied inside the LISA bandwidth, allowing for a direct measurement of the increase of eccentricity in the waveform of the system.  We find that such an exciting event could happen for an outer period of a few days or less, which could be the case if a binary system was stuck in a migration trap in disks around SMBHs~\cite{Bellovary:2015ifg}.

 % The possibility of observing such an exciting event calls for the development of waveform templates incorporating the effect of precession resonances, which is beyond the scope of this paper. Instead, we will now develop a semi-analytical toy model in order to gain physical insight on the evolution of $e$ at the resonance.

%TODO : check all order of magnitude in AU, especially the one for hubble time

\section{A semi-analytical model of the resonances} \label{sec:model_resonances}

In order to gain a physical understanding of the phenomenon at play when the resonances conditions are met, and to map efficiently the large parameter space, in this Section we will work out a simplified set of equations describing the evolution of the eccentricity of the inner binary at resonance. To this aim, we will make several simplifying assumptions in order to put the LPE~\eqref{eq:dot_a}-\eqref{eq:dot_Omega} in a tractable form:

(i) First, we will carry out our computations at lowest order in the eccentricity $e$, since it can be seen from the numerical solution that $e$ never reaches extreme values. We will also assume that $e$ is always sufficiently small so that its influence on radiation-reaction is negligible. This means that the semimajor axis $a$ can be expressed as
\begin{equation}\label{eq:at}
a(t) = a_0 \bigg( 1 - \frac{t}{t_\mathrm{RR}} \bigg)^{1/4} \; , \quad t_\mathrm{RR} = \frac{5 a_0^4}{256 G_N^3 m^3 \nu} \; ,
\end{equation}
where $a_0$ is the value of $a$ at $t=0$.
Radiation-reaction is essential in bringing the binary to a resonance, so we cannot ignore its effect. However, since $t_\mathrm{RR} \gg t_\mathrm{KL} \gg t_\mathrm{PN}$, we can always assume that the resonance takes places on timescales much shorter than $t_\mathrm{RR}$ and expand the power-law in Eq.~\eqref{eq:at} accordingly.

(ii) Second, we will place ourselves in the vicinity of the lowest-order resonance, defined by $\dot \omega = n_3/2$. Since, in the PN regime, the major contribution to $\dot \omega$ is given by the PN precession (Eq.\eqref{eq:omegadotPN}),
 this means that there exists a relation between the semimajor axis of the two orbits: 
\begin{equation} \label{eq:ExactResonanceCondition}
a_3 = a_0 \left( \frac{M a_0^2}{36 G_N^2 m^3} \right)^{1/3} \; .
\end{equation}
where we have chosen the initial time so that $t=0$ corresponds to the exact resonance $\dot \omega = n_3/2$ with $e=0$. 
Thus, at the resonance the KL timescale is
\begin{equation}
t_\mathrm{KL} = \frac{a_0 M}{36 m_3} \bigg( \frac{a_0}{G_N m} \bigg)^{5/2} \Rightarrow \frac{t_\mathrm{KL}}{t_\mathrm{PN}} = \frac{M a_0}{12 G_N m m_3 } \gg 1 \; ,
\end{equation}
which means that the PN timescale is always the shortest at the resonance, as expected.

(iii) Third, we will neglect the variation of the outer orbit planetary elements during the resonance. Indeed, it can be checked by using the LPE for the outer orbit that, at resonance, the timescale for the variation of the outer orbit elements is
\begin{equation}
t_\mathrm{out} \sim \frac{a_0}{\nu} \bigg( \frac{a_0}{G_N m} \bigg)^{17/6} \bigg( \frac{M}{m} \bigg)^{2/3}
\end{equation}
so that $t_\mathrm{out} \gg t_\mathrm{KL} \gg t_\mathrm{PN}$. This allows us to choose $\iota_3=0$ so that $\iota = \iota_\mathrm{tot}$ is the relative inclination between the two orbits.
%TODO : caution against the fact that outer elements vary in order to maintain H constant !

(iv) Fourth, we will collect in the LPE only the terms which are constant or in resonance, given that the other terms oscillate quickly in time and will average out. This is done by expanding the outer orbit's variables in the quadrupolar Hamiltonian as a Fourier series in time,
and then keeping in the Hamiltonian only terms which are constant or proportional to $\cos (2 \omega - n_3 t)$, $\sin (2 \omega - n_3 t)$, throwing away all other trigonometric functions of $\omega$ and $n_3 t$. We show in Appendix~\ref{sec:quad} the relevant formulaes for carrying out this Fourier series expansion, see Eq.~\eqref{eq:FourierSeries}.

Under this last simplifying assumption, the Hamiltonian of the inner binary becomes:
\begin{align}\label{eq:averaged_H}
\begin{split}
\frac{\mathcal{H}}{\mu} &= -3 \frac{G_N^2 m^2}{a^2 \sqrt{1-e^2}} - \frac{G_N a^2 m_3}{16 a_3^3} \bigg[ \frac{2+3e^2}{(1-e_3^2)^{3/2}} \big( 3 \cos^2 \iota-1\big) \\
& + 15 e^2 \big(f_1 \cos \psi + f_2 \sin \psi \big) \bigg]
\end{split}
\end{align}
where $\psi = 2 \omega - n_3 t$ is the resonant angle, and we did not simplify $\sqrt{1-e^2} \simeq 1$ in the first term of $\mathcal{H}$ (corresponding to the PN precession) since its magnitude is much greater than the second (quadrupolar) term, so that even a small $e$ could give a non-negligible contribution to the Hamiltonian.
The two angular factor $f_1$ and $f_2$ are defined by
\begin{align}
f_1 &= \frac{\mathsf{a}_1 - \tilde{\mathsf{a}}_1}{2} \big( 1+\cos^2 \iota\big) \cos 2 \omega_3 \nonumber \\
& - \mathsf{b}_1  \big( 1+\cos^2 \iota\big) \sin 2 \omega_3  + \frac{\mathsf{a}_1 + \tilde{\mathsf{a}}_1}{2} \sin^2 \iota \\
f_2 &= \cos \iota \big( 2\mathsf{b}_1 \cos 2 \omega_3 + (\mathsf{a}_1 - \tilde{\mathsf{a}}_1) \sin 2 \omega_3 \big) \; .
\end{align}
where $\mathsf{b}_1$, $\mathsf{a}_1$ and $\tilde{\mathsf{a}}_1$ are functions of the outer orbit eccentricity $e_3$ defined in Appendix~\ref{sec:quad}, Eqs~\eqref{eq:def_a1_Fourier}-\eqref{eq:def_b1_Fourier}.
Note that the terms in the first line of the Hamiltonian~\eqref{eq:averaged_H} correspond to the well-known orbit-averaged Hamiltonian (as given in e.g.~\cite{Naoz_2016}, setting to zero the $\cos \omega$ term as required by our assumption (iv)), while the second line contains the resonant terms. Furthermore, if $e_3 \rightarrow 0$, then $f_1, f_2 \rightarrow 0$, which means that the lowest-order resonance is suppressed in the case of a circular outer orbit. An explicit computation shows that the next resonance $\dot \omega = n_3$ would still persist in the case where $e_3=0$, although its effect on the eccentricity of the inner binary is generically weaker than the lowest-order resonance, as can be seen in Figure~\ref{fig1}. %Finally, notice that since $\mathsf{b}_1 \sim \mathsf{a}_1 - \tilde{\mathsf{a}}_1 \ll \mathsf{a}_1 + \tilde{\mathsf{a}}_1$ for generic $e_3$, one can approximate $f_2 \simeq 0$, $f_1 \simeq (\mathsf{a}_1 + \tilde{\mathsf{a}}_1) \sin^2 \iota/2$, which means that the strength of the resonance is maximized at $\iota = \pi/2$.

It is now quite easy to find the evolution of the inner planetary elements using the LPE~\eqref{eq:dot_a}-\eqref{eq:dot_Omega}. We find that $\dot \iota$ is suppressed by an eccentricity factor $e^2$, so that its variation is quite negligible. This is confirmed from our numerical solution. Thus, we will assume $\iota$ fixed in the following. The system then reduces to two differential equations on $e$ and $\psi$, which we write as
\begin{align}
\dot e &= \frac{15 e}{8 t_\mathrm{KL}} \big( f_1 \sin \psi - f_2 \cos \psi \big) \; , \label{eq:edot} \\
\dot \psi &= \frac{15}{4 t_\mathrm{KL}} \big( f_1 \cos \psi + f_2 \sin \psi \big) + \frac{2 e^2}{t_\mathrm{PN}} + \frac{t}{\tau^2}  \nonumber \\
&+ \frac{3(5 \cos^2 \iota -1)}{4 t_\mathrm{KL} (1-e_3^2)^{3/2}} \; , \label{eq:psidot}
\end{align}
where the new timescale $\tau$ is a mixture of RR and PN timescales:
\begin{align}
\tau &= \bigg(\frac{4 t_\mathrm{PN} t_\mathrm{RR}}{5} \bigg)^{1/2} = \frac{a_0}{8 \sqrt{3 \nu}} \bigg( \frac{a_0}{G_N m} \bigg)^{9/4} \label{eq:tau} \; .
\end{align} 
In order to obtain Eq.~\eqref{eq:psidot}, we have expanded the PN precession term for small eccentricities and small $t/t_\mathrm{RR}$, and we have neglected higher-order terms (this also means that we neglect the variation of $a$ in the quadrupolar Hamiltonian, i.e we set $a=a_0$ everywhere except in the PN term whose magnitude is the greatest).
% We have also chosen the initial time such that $t=0$ corresponds to the exact resonance with $e=0$, i.e
%\begin{equation} \label{eq:exactResonanceCondition}
%3 a_0 \bigg( \frac{G_N m}{a_0} \bigg)^{3/2}  = n_3
%\end{equation}
From the three timescales displayed in Eqs.~\eqref{eq:edot}-\eqref{eq:psidot}, the PN timescale $t_\mathrm{PN}$ is always the shortest one. On the other hand, the ratio of $t_\mathrm{KL}$ to $\tau$ is
\begin{equation}
\frac{t_\mathrm{KL}}{\tau} = \frac{16 M}{135 m_3} \sqrt{3 \nu}  \bigg( \frac{a_0}{G_N m} \bigg)^{1/4} \; ,
\end{equation}
so that it can be either larger or less than one, depending on the mass ratios and the relativistic parameter. For example, in the system considered in Figure~\ref{fig1}, the ratio $t_\mathrm{KL}/\tau$ is of order unity at the resonance. We will see later on that the precise value of this ratio determines the exponential growth of $e$. Finally, let us state once again that we will not neglect the first term proportional to $e^2$ in Eq.~\eqref{eq:psidot}, since $t_\mathrm{PN}$ is the shortest timescale and even a small eccentricity could give a term of order $1/t_\mathrm{KL}$ in~\eqref{eq:psidot}. We will indeed see in the following that this term plays a crucial role in the resonance. 

In order to bring the equations in a form which depends only on a few parameters, we will now neglect the coefficient $f_2$ in what follows. Indeed, notice that since $\mathsf{b}_1 \sim \mathsf{a}_1 - \tilde{\mathsf{a}}_1 \ll \mathsf{a}_1 + \tilde{\mathsf{a}}_1$ for generic $e_3$ (see Fig~\ref{fig:Fourier}), one can approximate $f_2 \simeq 0$, $f_1 \simeq (\mathsf{a}_1 + \tilde{\mathsf{a}}_1) \sin^2 \iota/2$ for angles $\iota$ which are not too small. This also means that the strength of the resonance is maximized at $\iota = \pi/2$. Introducing the dimensionless time $\tilde t = 15 f_1 t/8 t_\mathrm{KL}$, the system~\eqref{eq:edot}-\eqref{eq:psidot} can be brought in a form depending on three parameters only:
\begin{align}
e' &= e \sin \psi \; , \label{eq:edot_adim}\\
\psi' &= 2 \cos \psi + \alpha e^2 + \beta \tilde t + \gamma \label{eq:psidot_adim} \; ,
\end{align}
where a prime denotes differentiation with respect to $\tilde t$, and the coefficients $\alpha$, $\beta$ and $\gamma$ are given by 
\begin{align} \label{eq:alphaBetaGamma}
\begin{split}
\alpha &= \frac{4 M}{45 m_3 f_1} \frac{a_0}{G_N m} \; , \quad \beta = \frac{256 \nu M^2}{6075 m_3^2 f_1^2} \bigg(  \frac{a_0}{G_N m} \bigg)^{1/2} \; , \\
\gamma &= \frac{2(5 \cos^2 \iota -1)}{5 f_1 (1-e_3^2)^{3/2}} \; .
\end{split}
\end{align}

The parameter $\alpha \gg 1$ represents the competition between PN and KL timescales at resonance.  Furthermore, we will show later on that $\alpha$ sets the value of the maximum eccentricity $e$ that one can achieve at the resonance.
 $\beta \sim 1$ represents a mixture between radiation-reaction, PN and KL timescales. A small $\beta$ means that the radiation-reaction timescale is slow, so that the resonance will be quite effective at producing high eccentricities since the system spends a lot of time in resonance; on the other hand, a high $\beta$ will generically produce small eccentricities since the system crosses rather quickly the resonance condition when the radiation-reaction timescale is short. Finally, $\gamma \sim 1$ depends only on $\iota$ and the outer eccentricity $e_3$. %Furthermore, $\alpha$ sets the value of the maximum eccentricity $e$ that one can achieve at the resonance, via the formula $e_\mathrm{max}^2 = 2 \pi / \alpha$ derived in Eq.\eqref{eq:emax}.

\begin{figure}
\includegraphics[width=\columnwidth]{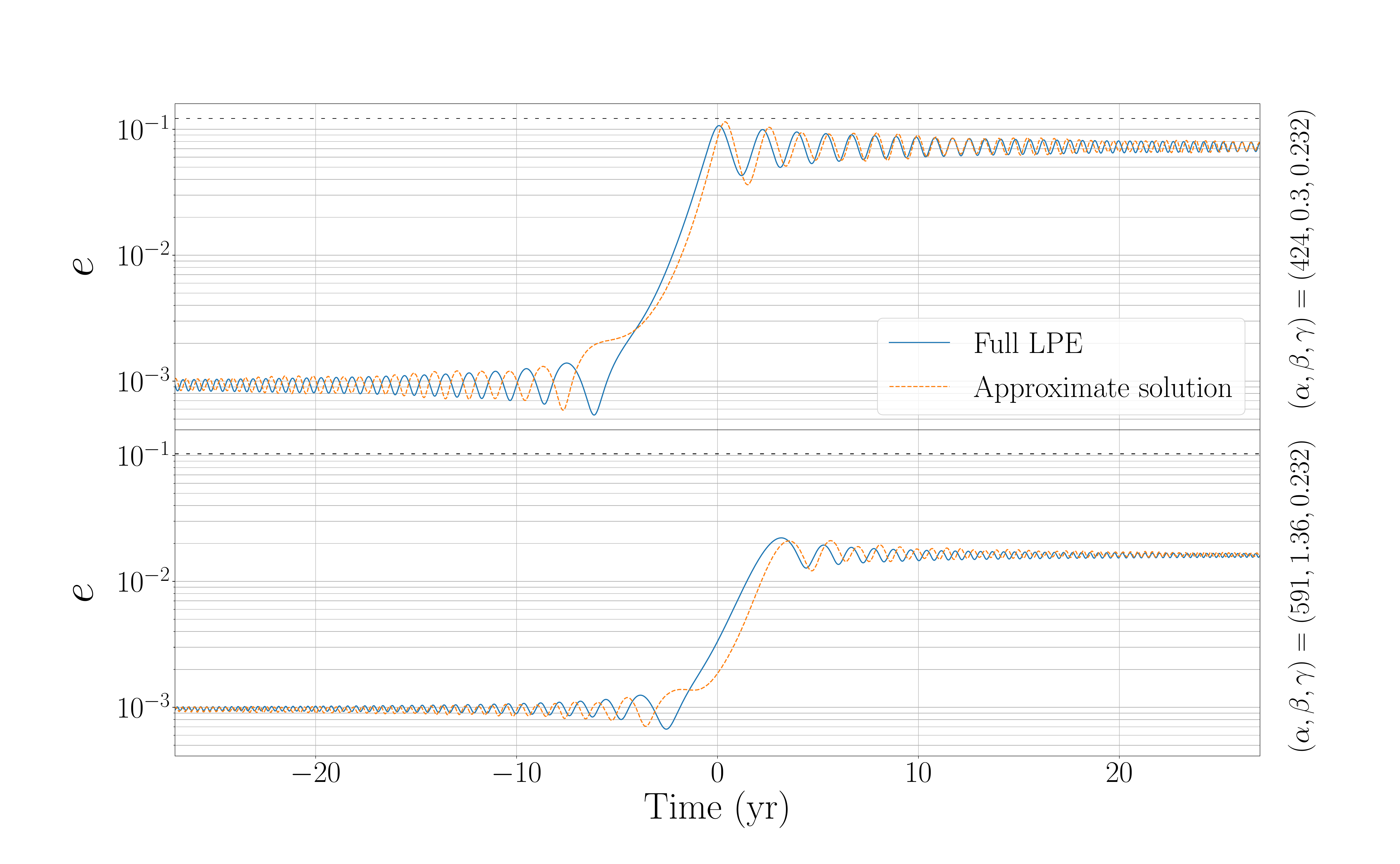}
\caption{\textit{Comparison between full and approximate solutions}: The upper panel shows a system with parameters $m_3=m=50 M_\odot$, $\nu=0.1$, $a_3=0.5$ AU, $e_3=0.7$, $\iota = 70 \degree$, $\omega_3=\omega=\Omega=0 \degree$ and initial eccentricity $e=10^{-3}$, corresponding to $(\alpha, \beta, \gamma) = (424, 0.3, 0.232)$. The lower panel has $m_3=30 M_\odot$, $\nu=0.25$ and the same other parameters, corresponding to $(\alpha, \beta, \gamma) = (591, 1.36, 0.232)$. In continuous blue is shown the solution from the full LPE, while in dashed orange is the solution from the simplified system~\eqref{eq:edot_adim}-\eqref{eq:psidot_adim}. The central time $t=0$ is chosen to correspond to the instant where $a=a_0$, where $a_0$ is defined in Eq.~\eqref{eq:ExactResonanceCondition}. The horizontal dahsed line correspond to the maximal possible eccentricity $e_\mathrm{max}^2=2\pi/\alpha$. The system from the upper panel reaches $e_\mathrm{max}$ while staying in resonance, whereas the one from the lower panel exits resonance before reaching $e_\mathrm{max}$.
}
\label{fig:comp}
%\textit{Comparison between full and approximate solutions}: 
\end{figure}

We were not able to find an analytic solution to the system~\eqref{eq:edot_adim}-\eqref{eq:psidot_adim}; however, it is now easy to build an understanding of the physical phenomenon at play. Indeed, if $\psi$ varies quickly in time, then the derivative of $e$ in Eq.\eqref{eq:edot_adim} will average out and $e$ will just feature small oscillations. On the other hand, when $\psi$ varies slowly i.e. when $\tilde t$ is such that $\psi'=0$ in Eq.~\eqref{eq:psidot_adim}, then the $\psi$-dependent term in Eq.\eqref{eq:edot_adim} can be considered as constant and $e$ undergoes exponential growth. This growth cannot last forever since the increase of eccentricity will move the system out of the resonance condition $\dot \omega = n_3/2$ from the fact that $\dot \omega$ depends on $e$, cf Eq.~\eqref{eq:omegadotPN}. Equivalently, requiring that the second term in Eq.~\eqref{eq:psidot_adim} should add a maximum dephasing of $\pi$ to the phase $\psi$ (meaning that $e'$ changes sign), one gets that the maximum eccentricity that one can possibly reach at resonance is given by
\begin{equation} \label{eq:emax}
e_\mathrm{max}^2 = \frac{2 \pi}{\alpha}  \; ,
\end{equation}
%where we have also maximized the result over the angle at resonance $\psi_m$. 
From this result, we can see that the resonance is efficient for smaller $\alpha$, e.g. when $m_3 \gtrsim m$ and when $f_1$ is maximal,  which corresponds to a perpendicular inclination and a large $e_3$. However, the binary does not always reach $e_\mathrm{max}$ during resonance if the radiation-reaction timescale is fast, as Figure~\ref{fig:comp} illustrates. 

These remarks allow for a qualitative, but not quantitative, understanding of the resonance.
In the following, we will then adopt a semi-analytic approach by solving the simplified system~\eqref{eq:edot_adim}-\eqref{eq:psidot_adim} numerically, which is a lot faster than integrating the full LPE.
In Figure~\ref{fig:comp} we show the comparison between $e$ obtained from the LPE and the simplified system~\eqref{eq:edot_adim}-\eqref{eq:psidot_adim}. One can see that our model is quite effective at reproducing the LPE solution, 
even if the exact period of the oscillations is slightly different in the two cases. 
%apart from a dephasing induced by our different approximations.
%a beating pattern which is not captured within our approximations.

\begin{figure}
\includegraphics[width=\columnwidth]{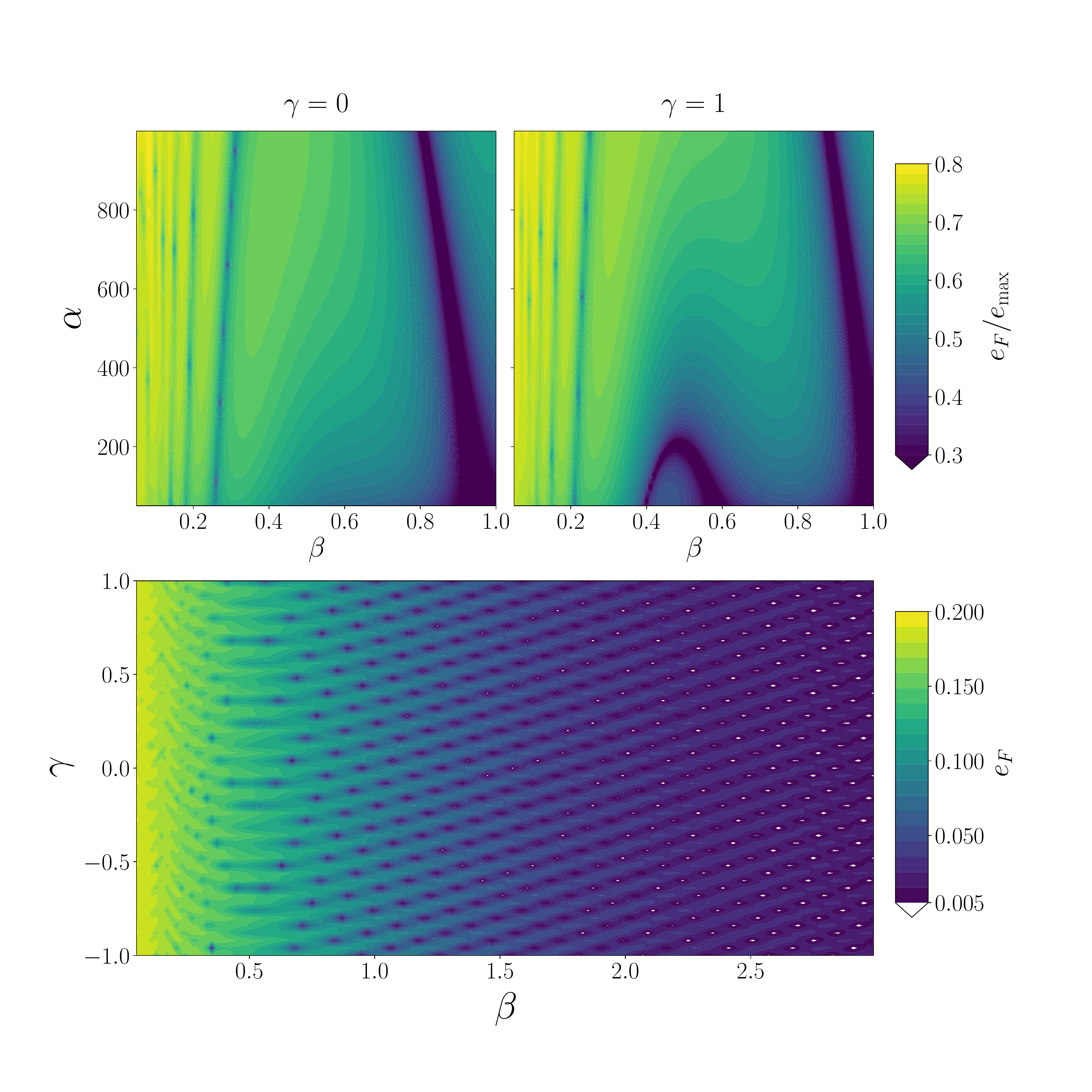}
\caption{\textit{Final eccentricity after resonance}: The final eccentricity $e_F$ is extracted by solving the simplified system~\eqref{eq:edot_adim}-\eqref{eq:psidot_adim} with initial values $\psi = 0$, $e = 5 \times 10^{-3}$ at $\tilde t = -100$.
 The two upper color plots show $e_F$ normalized to the maximal eccentricity $e_\mathrm{max}$ in the $(\alpha, \beta)$ plane for $\beta <1$ and for the two values $\gamma=0,1$. There appears to be some bands where the resonance is uneffective at enhancing the eccentricity. Changing $\gamma$ modifies the location of the bands but has little effect outside them. The lower color plot shows $e_F$ in the $(\gamma,\beta)$ plane for $\alpha = 100$ (corresponding to $e_\mathrm{max} = 0.25$). Notice that when $\beta \gtrsim 1$ the eccentricity can actually be decreased by the resonance (this is shown in white color).
}
\label{fig:colorPlot}
\end{figure}

We then use our simplified system~\eqref{eq:edot_adim}-\eqref{eq:psidot_adim} to obtain the final eccentricity $e_F$ reached after resonance, varying $\alpha$, $\beta$ and $\gamma$. One can observe two different behaviours depending on $\beta$. If $\beta \lesssim 1$, the resonance is effective and $e_F$ is mostly limited by the maximal eccentricity $e_\mathrm{max}^2 = 2 \pi / \alpha$.
On the other hand, if $\beta \gtrsim 1$, the system will never reach $e_\mathrm{max}$ and $\alpha$ has very little influence on $e_F$. We show a color plot of $e_F$ in both $(\alpha, \beta)$ and $(\beta, \gamma)$ plane in Fig~\ref{fig:colorPlot}, which unveils bands corresponding to larger or smaller $e_F$ values. We also find that $e_F$ can be decreased with respect to the initial value of $e$ only when $\beta \gtrsim 1$. 
%This last point can be physically understood from the fact that, if the RR timescale is slow, the resonant angle $\psi_m$ can always evolve towards a value which will lead to an exponential increase of $e$ in Eq.~\eqref{eq:e_crude}; on the other hand, if the RR timescale is quick, the resonant angle does not evolve during the resonance and can lead to an exponential decrease of $e$ for a range of values of $\psi_m$. 
A decrease of eccentricity is also an interesting phenomenon in itself, since it would mean for example that a system with $e$ brought to large values by the standard KL mechanism could see its eccentricity decreased by the resonance, which pushes its merger time towards higher values. %However, a detailed investigation of the effect of resonances on the eccentricity distribution of BHB is beyond the scope of this \textit{Letter}.

%furthermore $\gamma$ has little influence on $e_F$ in this case. We show in Fig~\ref{fig:colorPlot} a color plot of $e_F$ in the $(\alpha, \beta)$ plane, for two values of $\gamma$.

%~\footnote{On the other hand, we will assume that terms of order $\mathcal{O}(e^4)/t_\mathrm{PN}$ are indeed negligible}

\section{Conclusions}

The relativistic resonance phenomenon described for the first time in this article may drastically modify the parameters of a binary system perturbed by a distant mass. Indeed, a binary in an initial quasi-circular orbit undergoing a precession resonance may be brought to a new state whose eccentricity is potentially measurable by low-frequency GW detectors as LISA. When using the orbital eccentricity to discriminate against formation channels of BH, it will be important to take into account precession resonances. Thus, an important follow-up of the present work will be to include the effect of precession resonances into population studies which for the time being consider only the impact of the KL mechanism on BHB eccentricities~\cite{2018ApJ...856..140H,Nishizawa:2016eza,10.1093/mnras/stu039}. Another exciting prospect is the possibility of observing a precession resonance directly in the waveform of a BHB in the LISA band, which calls for new waveforms templates incorporating the effect of such a phenomenon. This could ultimately lead to a direct measurement of the parameters of the outer orbit from the inner binary waveform, thus giving important information on the properties of such three-body systems.

\section*{Acknowledgments}

I would like to thank an anonymous referee for pertinent remarks which led to the addition of Appendix~\ref{sec:higher}.
This research was partly supported by the Italian MIUR under contract 2017FMJFMW (PRIN2017).

\appendix

\section{Quadrupolar Hamiltonian} \label{sec:quad}

\begin{figure}
\includegraphics[width=\columnwidth]{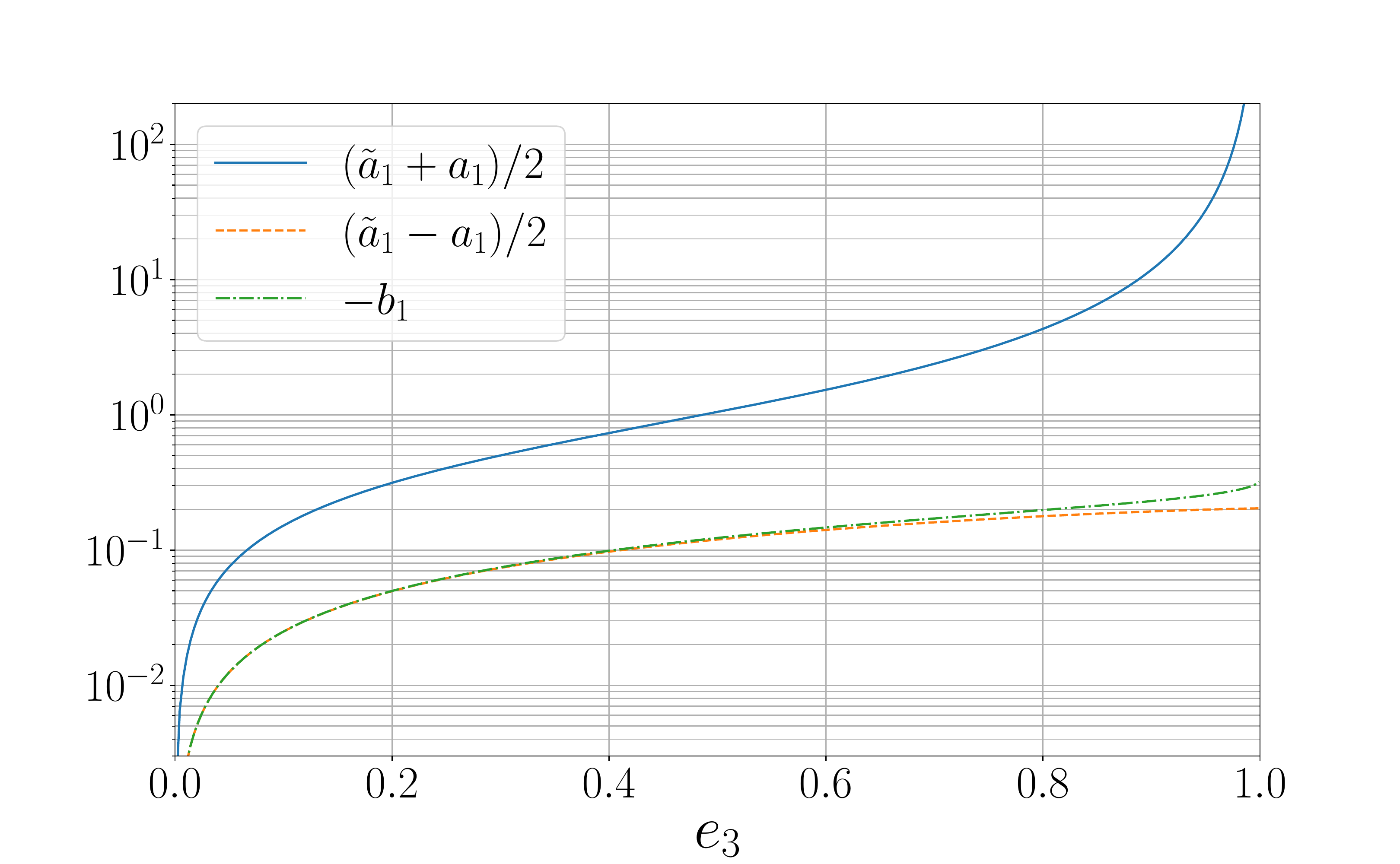}
\caption{ Plot of the three Fourier coefficients defined in Eqs.~\eqref{eq:def_a1_Fourier}-\eqref{eq:def_b1_Fourier}, showing the hierarchy $\mathsf{b}_1 \sim \mathsf{a}_1 - \tilde{\mathsf{a}}_1 \ll \mathsf{a}_1 + \tilde{\mathsf{a}}_1$ for any $e_3$. Notice that $\mathsf{a}_1 + \tilde{\mathsf{a}}_1$ diverges when $e_3 \rightarrow 1$, so that the effect of the resonance are maximized for large outer eccentricites $e_3$. However, if $e_3$ is too close to $1$, one would get to a point where the resonance is so powerful that our approximation (i) mentioned in Section~\ref{sec:model_resonances} ($e \ll 1$) would break down, signaling that our simplified system would not acurately reproduce the solution of the LPE.
}
\label{fig:Fourier}
\end{figure}

In the quadrupolar approximation, the Hamiltonian $\mathcal{H}_\mathrm{quad}$ of the three-body system used in Eq.~\eqref{eq:Hamiltonian} is given by expanding the full three-body Newtonian Hamiltonian in the center-of-mass frame of the inner binary, with result~\cite{Kuntz:2021ohi}
\begin{equation}\label{eq:Hquad}
\mathcal{H}_\mathrm{quad} = - \frac{3 G_N m_3}{2 R^3} Q_{ij} N^i N^j \; ,
\end{equation}
where $\mathbf{R}$ is the radius vector of the outer orbit, $R$ its norm and $\mathbf{N} = \mathbf{R}/R$ its unit vector. The traceless quadrupole moment of the inner binary, averaged over one orbit, is~\cite{Kuntz:2021ohi}
\begin{equation}
Q^{ij} = \frac{\mu a^2}{2} \bigg[ (1+4e^2) \alpha^i \alpha^j + (1-e^2) \beta^i \beta^j - \frac{2+3e^2}{3} \delta^{ij}  \bigg] \; ,
\end{equation}
where $\bm{\alpha}$ (resp. $\bm{\gamma}$) is the unit vector directed towards the perihelion (resp. angular momentum), and $\bm \beta = \bm \gamma \times \bm \alpha$. In terms of the osculating elements, the expression of these vectors is
\begin{align} \label{eq:param_e_l}
\begin{split}
\bm{\alpha} &= R_z(\Omega) R_x(\iota) R_z(\omega)  \mathbf{u}_x \; , \\ \quad \bm{\beta} &= R_z(\Omega) R_x(\iota) R_z(\omega)  \mathbf{u}_y  \; , \\ \quad \bm{\gamma} &= R_z(\Omega) R_x(\iota) R_z(\omega)  \mathbf{u}_z \; ,
\end{split}
\end{align}
where $ \mathbf{u}_x$, $\mathbf{u}_y$, $\mathbf{u}_z$ are the Cartesian basis vectors. In a similar way, one can define the unit vectors parametrizing the orientation of the outer orbit, which we denote by $\bm \alpha_3$, $\bm \beta_3$ and $\bm \gamma_3$.
We can finally give the expression of the radius vector of the outer orbit in terms of its osculating elements:
\begin{align}
R &= a_3 (1 - e_3 \cos \eta_3) \; ,  \label{eq:def_R}\\
\bm N &= \frac{\cos \eta_3 - e_3}{1 - e_3 \cos \eta_3} \bm \alpha_3 + \sqrt{1-e_3^2} \frac{\sin \eta_3}{1 - e_3 \cos \eta_3} \bm \beta_3  \label{eq:def_N}\; ,
\end{align}
where $\eta_3$ is the eccentric anomaly of the outer orbit defined by $\eta_3 - e_3 \sin \eta_3 = n_3 t$. In the following, it will be useful to expand the quadrupolar Hamiltonian~\ref{eq:Hquad} as a Fourier series in time, using
\begin{align} \label{eq:FourierSeries}
\begin{split}
\frac{N^i N^j}{R^3} &= \frac{1}{a_3^3} \bigg[ \big( \mathsf{a}_0 + \mathsf{a}_1 \cos n_3 t \big) \alpha_3^i \alpha_3^j + \big( \tilde{ \mathsf{a}}_0 + \tilde{\mathsf{a}}_1 \cos n_3 t \big) \beta_3^i \beta_3^j \\
&+ \mathsf{b}_1 \sin n_3 t \big( \alpha_3^i \beta_3^j + \alpha_3^j \beta_3^i \big) \bigg] \; ,
\end{split}
\end{align}
The eccentricity-dependent coefficients in the Fourier series~\eqref{eq:FourierSeries} are given by
\begin{align}
\mathsf{a}_0 &= \tilde{\mathsf{a}}_0 = \frac{1}{2 (1-e_3^2)^{3/2}} \; , \label{eq:def_a0_Fourier} \\
\mathsf{a}_1 &= \frac{1}{\pi} \int_{-\pi}^\pi \frac{(\cos \eta - e_3)^2}{(1-e_3 \cos \eta)^4} \cos (\eta - e_3 \sin \eta) \mathrm{d}\eta \; ,  \label{eq:def_a1_Fourier}\\
\tilde{\mathsf{a}}_1 &= \frac{1}{\pi} \int_{-\pi}^\pi \frac{(1 - e_3^2) \sin^2 \eta}{(1-e_3 \cos \eta)^4} \cos (\eta - e_3 \sin \eta) \mathrm{d}\eta \; , \\
\mathsf{b}_1 &= \frac{1}{\pi} \int_{-\pi}^\pi \frac{\sqrt{1-e_3^2} (\cos \eta - e_3) \sin \eta}{(1-e_3 \cos \eta)^4} \sin (\eta - e_3 \sin \eta) \mathrm{d}\eta \label{eq:def_b1_Fourier} \; ,
\end{align}
and they are plotted in Figure~\ref{fig:Fourier}. We have ignored the higher order coefficients in the Fourier series~\eqref{eq:FourierSeries} because they give terms relevant only for higher-order resonances, $\dot \omega = p n_3 /2$ with $p>1$. Note that the coefficients $\mathsf{a}_0$, $\tilde{\mathsf{a}}_0$ correspond to the usual orbit-averaged Hamiltonian, while the other coefficients multiply terms oscillating at the frequency of the outer orbit.

The equations of motion stemming from the Hamiltonian~\eqref{eq:Hamiltonian}, and which describe the evolution of planetary elements of both orbits on long timescales, are called the Lagrange Planetary Equations (LPE), given by
\begin{align}
\dot a &= - \sqrt{\frac{4a}{G_N m}} \frac{\partial  \tilde{\mathcal{H}}}{\partial u} \; , \label{eq:dot_a} \\
\dot e &=  \sqrt{\frac{1-e^2}{G_N m a e^2}} \frac{\partial  \tilde{\mathcal{H}}}{\partial \omega} - \frac{1-e^2}{\sqrt{G_N m a}e} \frac{\partial  \tilde{\mathcal{H}}}{\partial u} \; , \label{eq:dot_e} \\
\dot \iota& =  \frac{1}{\sqrt{G_N m a(1-e^2)} \sin \iota} \frac{\partial  \tilde{\mathcal{H}}}{\partial \Omega} \nonumber  \\
&- \frac{\cos \iota}{\sqrt{G_N m a(1-e^2)} \sin \iota} \frac{\partial  \tilde{\mathcal{H}}}{\partial \omega} \; , \\
\dot u &= \sqrt{\frac{4a}{G_N m}} \frac{\partial  \tilde{\mathcal{H}}}{\partial a} + \frac{1-e^2}{\sqrt{G_N m a}e} \frac{\partial  \tilde{\mathcal{H}}}{\partial e} \; , \label{eq:dot_M} \\
\dot \omega &= - \sqrt{\frac{1-e^2}{G_N m a e^2}} \frac{\partial \tilde{ \mathcal{H}}}{\partial e} + \frac{\cos \iota}{ \sqrt{G_N m a(1-e^2)} \sin \iota} \frac{\partial \tilde{ \mathcal{H}}}{\partial \iota} \; , \label{eq:dot_omega} \\
\dot \Omega &=  -\frac{1}{\sqrt{G_N m a(1-e^2)} \sin \iota} \frac{\partial \tilde{\mathcal{H}}}{\partial \iota} \; , \label{eq:dot_Omega}
\end{align} %\sqrt{\frac{G_N m}{a^3}} + 
where $\tilde{\mathcal{H}} = \mathcal{H}/ \mu$ (see e.g.~\cite{valtonen_karttunen_2006} for a derivation; in particular, note that, while they are often used with orbit-averaged Hamiltonian, the LPE describe the evolution of a three-body system without any need for averaging. The presence of terms oscillating at the frequency of the outer orbit induces variations of the orbital elements on the outer binary timescale as in e.g.~\cite{Luo_2016}, however they are of too small magnitude to be seen in the plots of the main text). The equations for the outer orbit are the same provided one replaces the inner parameters by the outer ones. Furthermore, in the LPE one can use the elimination of nodes which means replacing $\Omega_3 \rightarrow \Omega+\pi$~\cite{10.1093/mnras/stt302,Naoz_2016}. Finally, we will also add to the LPE the effect of radiation-reaction on the inner orbit~\cite{PhysRev.131.435, Maggiore:1900zz}:
\begin{align}
\left. \frac{\mathrm{d}a}{\mathrm{d}t} \right\vert_\mathrm{RR} &= - \frac{64}{5} \frac{G_N^3 m^3 \nu}{a^3 (1-e^2)^{7/2}} \bigg(1 + \frac{73}{24}e^2 + \frac{37}{96} e^4 \bigg) \; , \\
\left. \frac{\mathrm{d}e}{\mathrm{d}t} \right\vert_\mathrm{RR} &= - \frac{304 e}{15} \frac{G_N^3 m^3 \nu}{a^4 (1-e^2)^{5/2}} \bigg(1 + \frac{121}{304}e^2 \bigg) \; ,
\end{align}
where $\nu = \mu/m$ is the symmetric mass ratio of the inner binary, and the evolution of other planetary elements is left unchanged by radiation-reaction.

Due to their complicated form, we will not display the final equations obtained by plugging the Hamiltonian~\eqref{eq:Hamiltonian} into the LPE~\eqref{eq:dot_a}-\eqref{eq:dot_Omega}. Instead, we will present a simplified version of them in Section~\ref{sec:model_resonances}.

\section{Higher-order terms in the Hamiltonian} \label{sec:higher}

\begin{figure}
\includegraphics[width=\columnwidth]{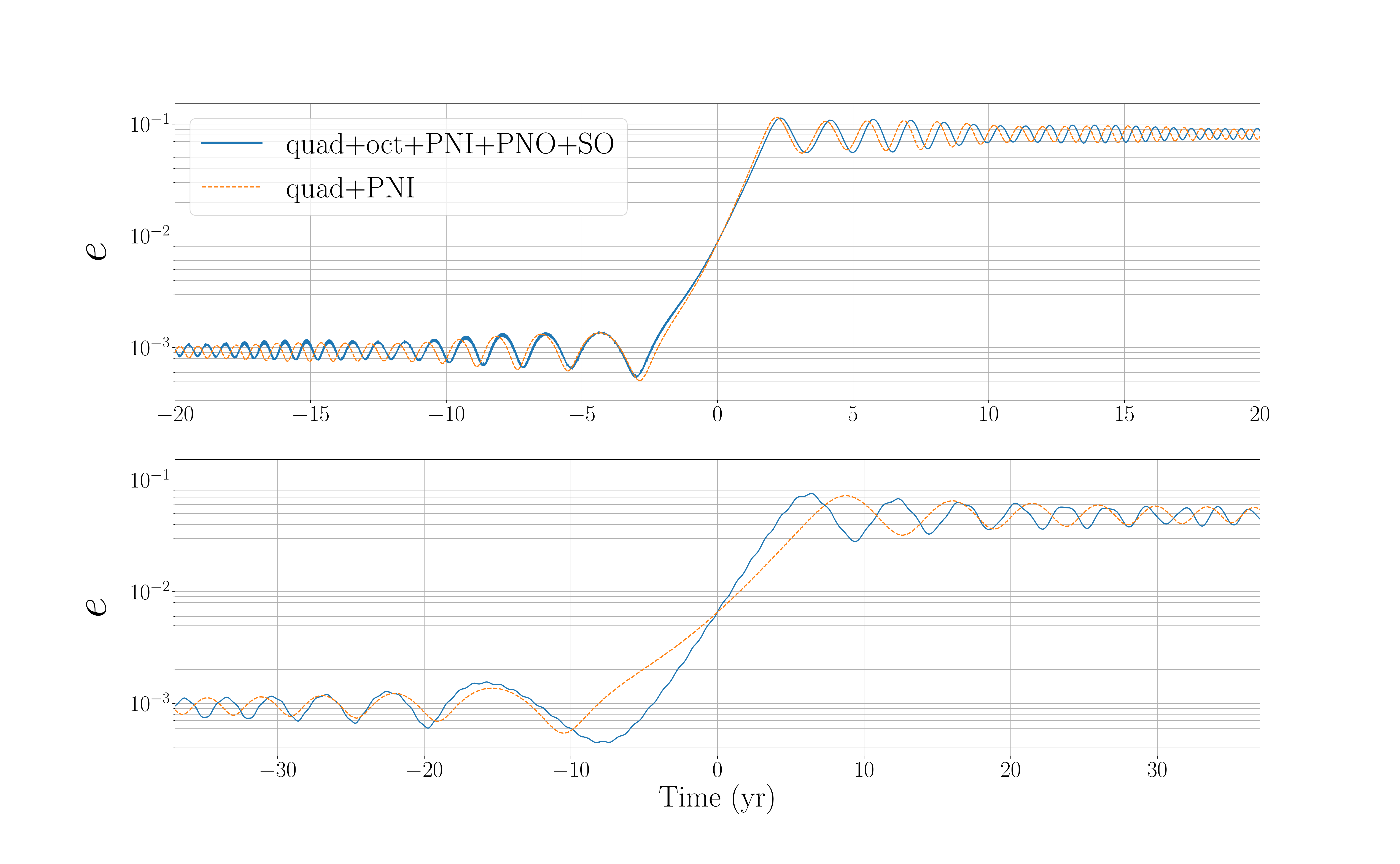}
\caption{\textit{Effect of higher-order terms on the resonance}: Eccentricity around the lowest-order resonance $\dot \omega = n_3/2$ for two model systems. The continuous blue curve corresponds to solving the LPE~\eqref{eq:dot_a}-\eqref{eq:dot_Omega} with an Hamiltonian composed of Eq.~\eqref{eq:Hamiltonian} plus the three higher-order terms in Eqs.~\eqref{eq:H_epsilon3}-\eqref{eq:H_v2epsilon3/2}, while the dashed orange curve corresponds to using the Hamiltonian~\eqref{eq:Hamiltonian} only as is done in the main text.
 The first system has the following parameters: $m=m_3=50 M_\odot$, $\nu = 0.1$,  $a_3=0.5$ AU, $e_3=0.7$, $\iota = 70 \degree$, $\omega_3=\omega=\Omega=0 \degree$ and initial eccentricity $e=10^{-3}$. The second system has parameters $m=50 M_\odot$, $m_3 = 4 \times 10^6 M_\odot$, $\nu = 0.1$,  $a_3=30$ AU, $e_3=0.6$, $\iota = 60 \degree$, $\omega_3=\omega=\Omega=0 \degree$ and initial eccentricity $e=10^{-3}$. Time is centered so that $t=0$ corresponds to the middle of the resonance.
}
\label{fig:octupole}
\end{figure}

An important question concerning the validity of our results is whether a quadrupolar approximation at lowest PN order is sufficient to accurately describe the motion of three-body systems that we describe in this article. Indeed, it is known that octupolar terms can drastically alter the KL mechanism~\cite{Naoz_2016} or lead to another kind of precession resonance~\cite{Liu:2015wgi, Liu_2020}, while 1PN terms induced by the outer binary can be important when the third body is supermassive~\cite{PhysRevD.89.044043, Lim:2020cvm}.
Can the inclusion of higher-order terms in the Hamiltonian~\ref{eq:Hamiltonian} change the resonance behavior? In this Section, we will explicitely prove that this is not the case.
%our Hamiltonian~\ref{eq:Hamiltonian} is adapted to describe the evolution of the kind of three-body systems considered in this article with sufficient accuracy.

The rationale for neglecting octupolar as well as other PN terms (such as precession of the outer orbit) in the Hamiltonian~\ref{eq:Hamiltonian} is the following. Consider first the evolution of the system far from resonance. The inclusion of higher-order terms will definitely affect the parameters of the outer binary (such as its argument of perihelion or eccentricity), but what about the inner orbit? In all cases described in this article, we are deep in the regime where PN terms dominate over quadrupolar perturbations $t_\mathrm{PN} \ll t_\mathrm{KL}$. As explained in the main text, this has the consequence of averaging the multipolar interactions between inner and outer binaries so that KL oscillations are suppressed and the evolution of inner binary is similar as an isolated binary. What is true for the quadrupolar perturbation is also true for the octupole or PN-multipolar cross-terms (which are further suppressed compared to the quadrupole), so that we can expect the inner binary to be insensible to higher-order terms in the Hamiltonian in this case. 
%On the other hand, the outer binary is not necessarily dominated by PN precession, so that higher-order terms can change its parameters (such as argument of perihelion or eccentricity) far from resonances.

Consider next the situation when the system is close to resonance. In all cases examined in the main text, the duration of the resonance is a few times the KL timescale $t_\mathrm{KL}$. Consequently, all higher-order terms in the Hamiltonian which are further suppressed with respect to the quadrupole cannot accumulate on long timescales and give appreciable variations in the osculating elements. In other words, their effect can only be seen on timescales much longer than a few KL timescales which is the duration of the resonance. Thus, in this case also higher-order terms in the Hamiltonian cannot give appreciable changes to the evolution of the three-body system at resonance.

Another effect that one can imagine is that the \textit{location} of the resonances would be displaced by higher-order terms. Again, this proves to be a small modification. Consider for example the case where GR effects of the outer orbit are taken into account. Then, the frequency of the outer orbit (defined by the derivative of mean anomaly $\dot u_3$) receives PN corrections from the LPE~\eqref{eq:dot_M}. Thus, the resonance condition will slightly shift from the previous one $\dot \omega = n_3$. %In a similar way, the resonance parameters defined in Eq.~\eqref{eq:alphaBetaGamma} 

A last effect of octupolar terms is to trigger new resonances of the form $3 \dot \omega = p n_3$ ($p \in \mathbb{Z}$), while the quadrupolar resonances discussed in the main text are characterized by $2 \dot \omega = p n_3$. These higher-order resonances are of course of smaller magnitude and we will not discuss them here.

In order to prove our statements, we will consider the evolution of two model systems when taking into account higher-order terms in the Hamiltonian, namely the octupole, GR precession of the outer orbit, and coupling between the spins of inner and outer binaries. This corresponds to a Newtonian precision of order $\varepsilon^3$ and 1PN of order $v^2 \varepsilon^{3/2}$ in the power-counting rules presented in~\cite{Kuntz:2021ohi}. We refer the reader to~\cite{Kuntz:2021ohi} for a derivation of the corresponding Hamiltonian; note that, in order to be consistent with the results presented in this article, we will not average this Hamiltonian over the outer orbit timescale. Thus, the supplemetary terms that we consider are
\begin{widetext}
\begin{align}
\mathcal{H}_\mathrm{\varepsilon^3} &= -\frac{G_N m_3}{2R^4} \mathcal{O}^{ijk} \big(5 N_i N_j N_k - \delta_{ij} N_k - \delta_{ik} N_j - \delta_{jk} N_i \big) \label{eq:H_epsilon3} \; , \\
\mathcal{H}_{v^2 \varepsilon} &= - \frac{G_N^2 m m_3 (m+m_3)}{2a_3^2 (1-e_3 \cos \eta_3)^2} \bigg[ \frac{1-3 \nu_3}{4} (1+e_3 \cos \eta_3)^2 + (3 + \nu_3)(1+e_3 \cos \eta_3) + \nu_3 e_3^2 \frac{\sin^2 \eta_3}{1 - e_3\cos \eta_3} -1  \bigg] \label{eq:H_v2epsilon} \; , \\
\mathcal{H}_{v^2 \varepsilon^{3/2}} &= \frac{G_N \mu m_3 (4m +3m_3)}{2(m+m_3)a_3^3(1-e_3\cos \eta_3)^3} \sqrt{G_N m a(1-e^2)} \sqrt{G_N (m+m_3) a_3 (1-e_3^2)} \bm \gamma \cdot \bm \gamma_3 \; , \label{eq:H_v2epsilon3/2}
\end{align}
\end{widetext}
where the notations used in these equations have been introduced in Appendix~\ref{sec:quad}.
Let us comment on each of these Hamiltonians. $\mathcal{H}_\mathrm{\varepsilon^3}$ is the octupolar Hamiltonian, obtained after expanding Newton's potential in the center-of-mass frame of the inner binary. It depends on the octupole moment of the inner binary
$\mathcal{O}^{ijk} = \nu (m_2-m_1) \langle r^i r^j r^k \rangle $  ($\mathbf{r}$ being the radius vector of the inner orbit, and $\nu$ its symmetric mass ratio) averaged over one orbit, given by
\begin{align}
\begin{split}
\mathcal{O}^{ijk} &= \frac{5 \nu}{8} (m_1-m_2) \bigg[ (3+4e^2) \alpha^i \alpha^j \alpha^k \\
&+ (1-e^2) \big(\alpha^i \beta^j \beta^k + \beta^i \alpha^j \beta^k + \beta^i \beta^j \alpha^k \big) \bigg] \; ,
\end{split}
\end{align}
where $\bm \alpha$ and $\bm \beta$ have been defined in Eq.~\eqref{eq:param_e_l}. %Finally, $R$ and $\bm N$ have been defined in Eq.~\eqref{eq:def_R} and Eq.~\eqref{eq:def_N} respectively.

$\mathcal{H}_{v^2 \varepsilon}$ is the 1PN Einstein-Infeld-Hoffmann Hamiltonian of the outer orbit, responsible for its perihelion precession. It depends on the outer eccentric anomaly $\eta_3$ (defined by $\eta_3 - e_3 \sin \eta_3 = n_3 t$) and the outer symmetric mass ratio $\nu_3 = m m_3/(m+m_3)^2$. Averaging it over the outer orbit, we would find a 1PN Hamiltonian similar to the one describing perihelion precession of the inner orbit, $\langle \mathcal{H}_{v^2 \varepsilon} \rangle = -3 G_N^2 m m_3 (m+m_3)/a_3^2 \sqrt{1-e_3^2}$. However, as previously emphasized we will not average it over the outer orbit for consistency.

Finally, $\mathcal{H}_{v^2 \varepsilon^{3/2}}$ is the coupling between the angular momentums of inner and outer binaries, derived e.g. in~\cite{Kuntz:2021ohi}.

We now solve for the evolution of the three-body system following the LPE~\eqref{eq:dot_a}-\eqref{eq:dot_Omega}.  We present our results in Figure~\ref{fig:octupole} for two model systems, one in which the perturber is of the same mass than the binary system, and one in which the perturber is supermassive. The behavior of the inner binary system away from resonance is, as expected, very similar to an isolated binary since all multipolar terms in the Hamiltonian are 'averaged out' by the quick PN precession. Close to resonance, the supplementary terms in the Hamiltonian introduce some dephasing and slightly change the duration of the resonance. This is only apparent when the perturber is supermassive (bottom of Figure~\ref{fig:octupole}). However, note that the final eccentricity reached after resonance is barely affected by the supplementary terms in both cases.
This justifies our previous arguments and proves that using the minimal Hamiltonian~\eqref{eq:Hamiltonian} is sufficient for describing the effect of precession resonances in the regime $t_\mathrm{PN} \ll t_\mathrm{KL}$.

\bibliography{Resonances.bib}

\end{document}